\documentclass[preprint,12pt]{elsarticle}

\usepackage{amssymb}
\usepackage{subfigure}
\usepackage{graphicx}

\journal{Physica A}

\begin{document}

\begin{frontmatter}



\title{Temporal statistical analysis on human article creation patterns}


\author[mpustc]{Na Li}
\author[mpustc]{Han Yan}
\author[mpustc]{Wen-Yao Zhang}
\author[jiangyin]{Yu-Jian Li}
\ead{jinzhili@mail.ustc.edu.cn}
\author[jiangyin]{Zhen-Dong Xi}
\author[mpustc]{Bing-Hong Wang}

\address[mpustc]{Department of Modern Physics, University of Science and Technology of China, Hefei 230026, China}
\address[jiangyin]{China Satellite Maritime Tracking and Control Department, Jiangyin  214400, China}

\begin{abstract}
Inspired by previous works on human dynamics,
we collect the temporal statistics of the article creation by three Western scientists and an Eastern writer.
We investigate the distributions of the time intervals between the creations of every two consecutive articles.
All four time distributions are found to be deviate from the Poisson statistics,
 and show an approximate power-law distribution.
The power-law exponents are different with respect to individual, indicating that
the article creation pattern may be subjected to influence of many ingredients of the individual.

\end{abstract}

\begin{keyword}
Human dynamics \sep Article creation patterns \sep Power-law distribution

\end{keyword}

\end{frontmatter}


\section{Introduction}
\label{sec:intro}

Standing at the intersection of research interests of sociology, statistical physics, economics and many other disciplines,
the study of human behaviors has attracted lots of attention in recent years.
Nowadays, the main scientific interests are moving from qualitative hypotheses and conclusions
to more quantitative ones \cite{barabasi:2007ieee,zhou:2008sm}.
Traditionally, the pattern of individual behavior is thought to be a random process conforming to the Poisson law,
which leads to an exponential time interval distribution\cite{haigt:1967poisson}.
Under such assumption, the probability density function of the time interval $\tau$ between two
consecutive events has the form of $p(\tau)=\lambda e^{-\lambda\tau}$.\\

However, more and more evidences from the study of human behaviors reveal
an alternative situation in coherence with a non-Poisson process \cite{barabasi:2005nature,goh:2008epl},
 whose characteristic is that the events erupt within certain time intervals,
while these intervals are separated by long terms of inactiveness.
Those heavy-tailed distributions,
emerging in many situations including sending SMS \cite{candia:2008jpa},
replying mails \cite{li:2008cs,li:2008pa}, transportation \cite{han:2011pre},
web browsing \cite{dezso:2006pre,zhao:aps},
can be well approximated by a power-law formula $p(\tau)=\tau^{-\gamma}$.
These discoveries have started a new research direction in \emph{human dynamics}.
Besides performing empirical analysis, scientists have also proposed mathematical models to
explain the underlying mechanism. Barab\'{a}si \emph{et al.} have proposed
to explain the burst of human dynamics as a result of queuing process,
controlled by human decision making\cite{barabasi:2005nature,vaquez:2006pre}.
In Ref.\cite{vaquez:2006pre}, two universality classes for human dynamics are presented,
whose characteristic power-law exponents are 1 and 1.5.\\

These discoveries have great values in both scientific research and practical application.
Therefore, not only in physics and biology \cite{6,7,8}, but also in sociology and economics \cite{9,10},
human dynamics have a received wide attention.
Its commercial application is also attractive, e.g.,
 telephone service providers' improvement of resource distribution and price plans \cite{1},
  as well as urban retail trading \cite{2,3,4,5}, as brought up by Barab\'{a}si.\\

Empirical explorations at the nascent stage of research on human dynamics are critical.
Without these analysis we will lost our sight of a comprehensive scenario of the real human activity patterns.
In this paper, we carry out a study of the article creation of three western scientists and an eastern writer.
We choose Albert-L\'{a}szl\'{o}  Barab\'{a}si, Mark Newman and Harry Eugene Stanley
 as the representatives of the western scientists and Su Shi
as the representative of the eastern writers. Barab\'{a}si, Newman and Stanley are accomplished scientist,
and Su Shi is one of the most distinguished poets in Chinese history.
We make our choice to include people from both western and eastern culture environment, modern and ancient society, science and literature, so that the results are more comprehensive and convincing.
The data of the western scientists come from Science Citation Index Expanded (SCIE).
The data of Su Shi come from Ref.\cite{sushi}.\\

Our results show that the patterns of article creation obey the power-law distribution.
However, different individual holds different power-law exponent.
This reflect the fact that the article creation pattern can be affected by
personal characteristics, so the dynamics of creation activity may be complicated. \\

\section{Empirical results}\label{sec:res}

\begin{figure}[htbp]
\begin{center}
  \includegraphics[width=0.7\textwidth]{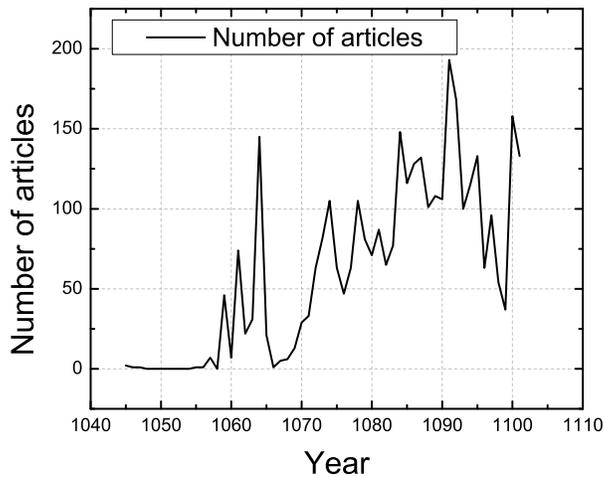}
  \caption{Historical record of the number of articles created by Su Shi each year.}\label{sushiyear}
\end{center}
\end{figure}

The data of the western scientists come from SCIE (Science Citation Index Expanded).
The data of Su Shi come from Ref.\cite{sushi}.
Fig.\ref{sushiyear} illustrates the historical record of the number of articles created by Su Shi each year.
The total numbers of records of each individual are listed in Table.\ref{record}.
The record numbers for the scientists are comparatively smaller,
so we choose a larger time span unit,
e.g., $15 d$ to $30 d$, to reduce the effect from statistical fluctuation. The record number for the Chinese poet is large enough for us to
choose a smaller time span unit.\\

\begin{table}
\begin{center}
\begin{tabular}{ c c  }
  \hline
  Individual & Total records  \\
  \hline
  \hline
  Stanley & $657$  \\
  Barab\'{a}si & $175$  \\
  Newman & $110$  \\
  Su Shi & $3213$ \\
  \hline
\end{tabular}
\end{center}
\caption{Detail of the records}\label{record}
\end{table}

\begin{figure}[htbp]
\centering
 \subfigure[Stanley]{
   \includegraphics[width=0.45\textwidth] {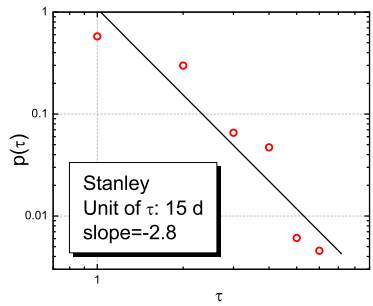}
   \label{power:stan}
 }
 \subfigure[Barab\'{a}si]{
   \includegraphics[width=0.45\textwidth] {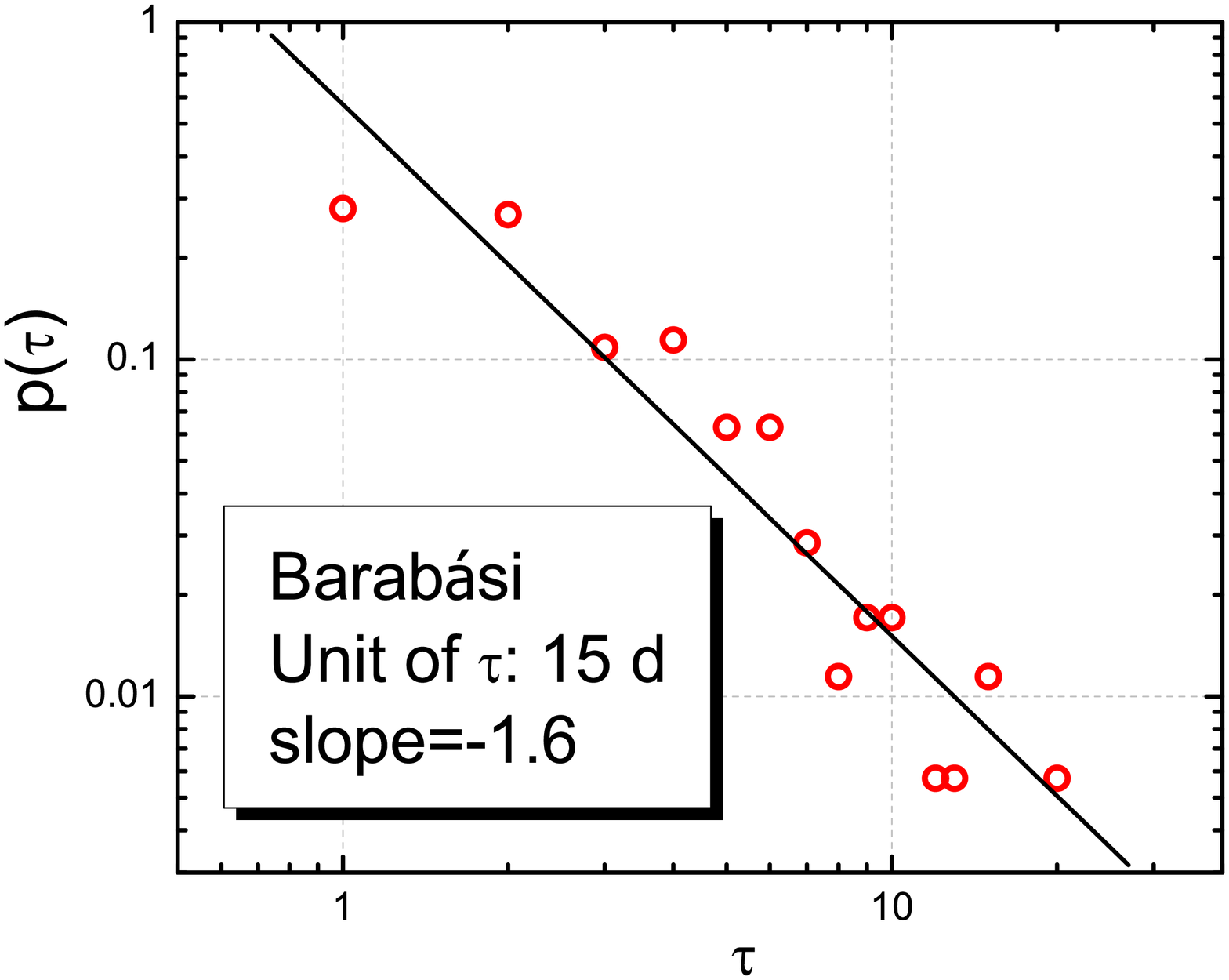}
   \label{power:bara}
 }
  \subfigure[Newman]{
   \includegraphics[width=0.45\textwidth] {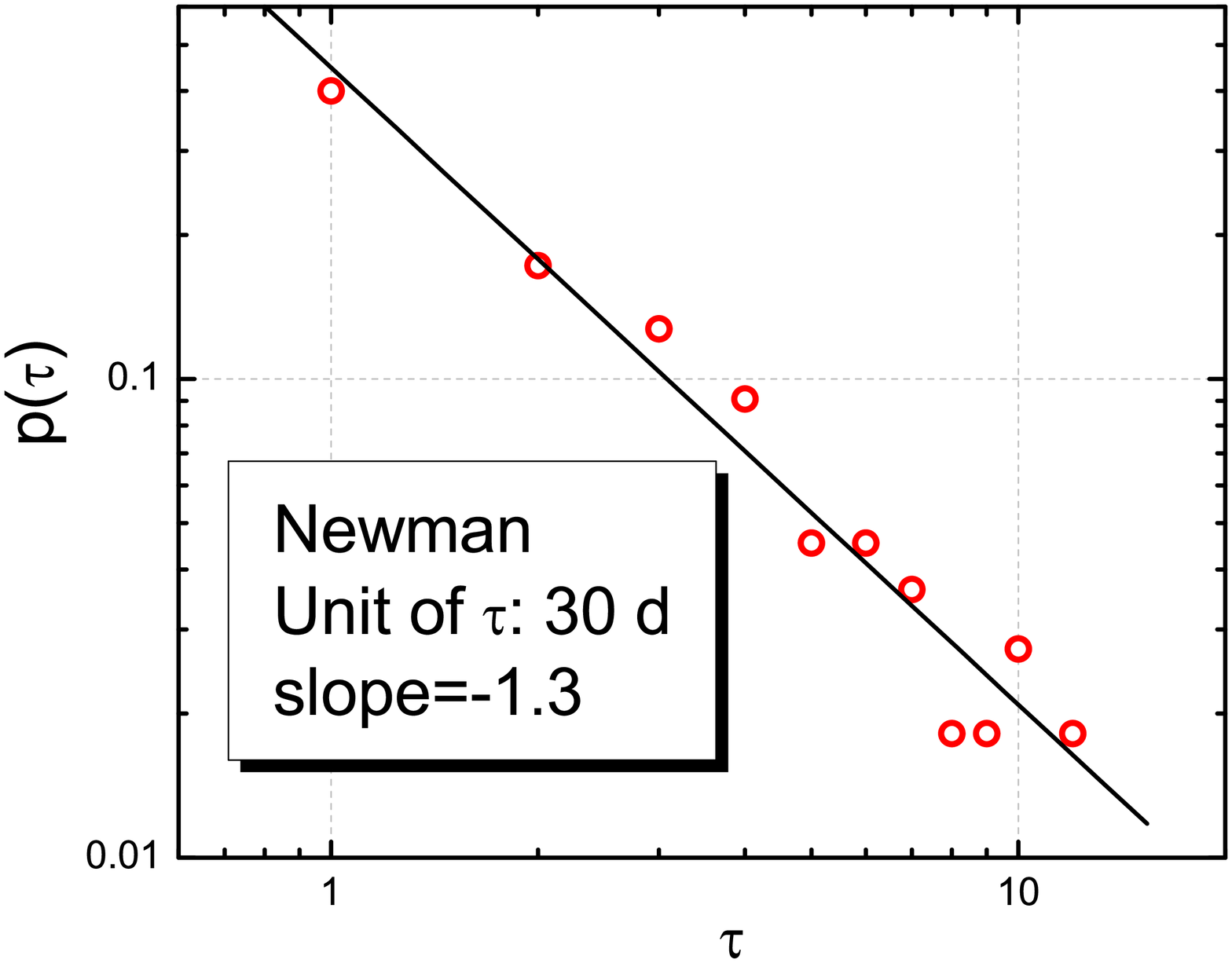}
   \label{power:newman}
 }
  \subfigure[Su Shi]{
   \includegraphics[width=0.45\textwidth] {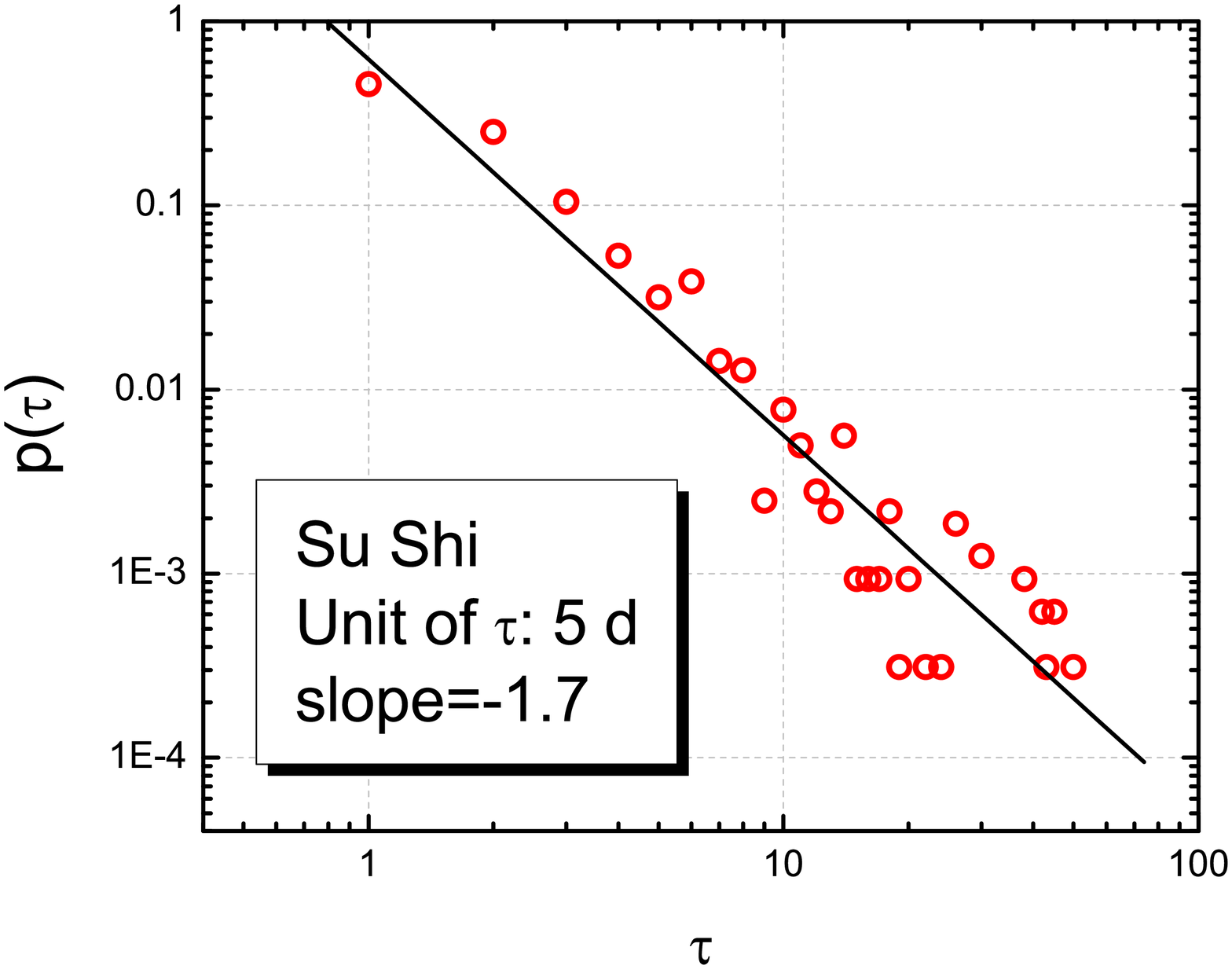}
   \label{power:su}
 }
\caption{The distribution of the article creation time intervals}\label{power}
\end{figure}

Next, we investigate the distribution of the time intervals between the
creations of every two consecutive articles. The time interval is defined as $\tau_{\omega}$.
As shown in Fig.\ref{power}, all the four distributions are well described by
the power law $p(\tau_{\omega})\sim \tau_{\omega}^{-\gamma_{\omega}}$.
the power-law exponents are $2.8$, $1.6$, $1.3$ and $1.7$
for Stanley, Barab\'{a}si, Newman and Su, respectively.

\section{Conclusion and discussion}\label{sec:con}

In this paper, the temporal statistics of the article creation time intervals of three Western scientists and a Chinese writer are empirically studied.
The distribution of the  time intervals of each individual can be well approximated by the power-law distribution. This non-trivial conclusion indicates that the human article creation activities share the same law with other activities like sending SMS \cite{candia:2008jpa},
replying mails \cite{li:2008cs,li:2008pa}, transportation \cite{han:2011pre},
and web browsing \cite{dezso:2006pre,zhao:aps}.
The underlying reason of the emergent of the power-law distribution is still open to question,
and further understanding of this issue is critical in human dynamics study.

However, the power-law exponents of each individual are different,
which reflects that there may be no universal pattern of human article creation activities.
There are numerous ingredients that may impact the creation activity pattern, e.g.,
memories \cite{vazquez:2007pa} and the interests \cite{han:2008njp}.
Further research of how these ingredients can affect the pattern of creation and other activities
are needed for a more comprehensive understanding of human dynamics.
Besides, we believe empirical studies of other creation activities,
 e.g., painting, composing, inventing are also needed to understand the pattern of creation.

\section*{Acknowledgements}
This work is funded by
the National Natural Science Foundation of China (Grant Nos.
9102402610975126, 10635040 and 11005001).



\newpage









\end{document}